\documentclass[twocolumn]{aastex631}

\usepackage{tablefootnote}

\begin{document}
\title{High-resolution mid-IR spectroscopy of SVS 13-A with EXES/SOFIA:\\
The surprisingly high CH$_3$OH/H$_2$O ratio in the planet-forming zone of a solar mass protostar.}

\author[0000-0002-6528-3836]{Curtis DeWitt}
\affiliation{Space Science Institute, 4765 Walnut St, Suite B, Boulder, CO, USA}

\author[0000-0001-5659-0140]{Marta De Simone}
\affiliation{ESO, Karl Schwarzchild Str. 2, D-85478 Garching bei München, Germany}
\affiliation{INAF, Osservatorio Astrofisico di Arcetri, Largo E. Fermi 5, I-50125, Firenze, Italy}

\author[0000-0001-9249-7082]{Eleonora Bianchi}
\affiliation{INAF, Osservatorio Astrofisico di Arcetri, Largo E. Fermi 5, I-50125, Firenze, Italy}

\author[0000-0001-9664-6292]{Cecilia Ceccarelli}
\affiliation{Univ. Grenoble Alpes, CNRS, IPAG, 38000 Grenoble, France}

\author[0000-0003-1514-3074]{Claudio Codella}
\affiliation{INAF, Osservatorio Astrofisico di Arcetri, Largo E. Fermi 5, I-50125, Firenze, Italy}
\affiliation{Univ. Grenoble Alpes, CNRS, IPAG, 38000 Grenoble, France} 

\author[0000-0002-7489-3142]{Sarah Nickerson}
\affiliation{Space Science and Astrobiology Division, NASA Ames Research Center, Moffet Field, CA 94035 USA}
\affiliation{Bay Area Environmental Research Institute, Moffett Field, CA 94035, USA}

\author[0000-0002-8030-7410]{Keeyoon Sung}
\affiliation{Jet Propulsion Laboratory, California Institute of Technology, Pasadena, CA 91011,USA}

\author[0000-0002-9637-4554]{Albert Rimola}
\affiliation{Universitat Autònoma de Barcelona: Bellaterra, Catalunya, Spain}

\author[0009-0000-7269-8418]{Vittorio Bariosco}
\affiliation{University of Turin: Turin, Piedmont, Italy}

\author[0000-0001-8886-9832]{Piero Uliengo}
\affiliation{University of Turin: Turin, Piedmont, Italy}

\author[0000-0001-9920-7391]{Naseem Rangwala}
\affiliation{Space Science and Astrobiology Division, NASA Ames Research Center, Moffet Field, CA 94035 USA}

\begin{abstract}

Water and methanol are key components of interstellar ices and gas in star- and planet-forming regions, but direct observations of water in low-mass protostars are challenging due to atmospheric absorption. We present high-resolution (R = 70,500) mid-infrared spectroscopy of the Class I protostar SVS13-A with EXES on board SOFIA at 26 $\mu$m, targeting both H$_2$O and CH$_3$OH absorption lines. Several lines of each species are detected, tracing warm gas with rotational temperatures of $\sim$140--170 K. Remarkably, the methanol column density is a factor of $\sim$4 higher than that of water, well above typical interstellar ice ratios ($<$10\%). Comparison with previous millimeter observations indicates that absorption and emission probe distinct regions, with the mid-IR lines likely tracing cooler gas along the line of sight. The surprising observed CH$_3$OH/H$_2$O ratio may reflect selective sublimation due to the distribution of binding energies or ice stratification in the inner envelope. These observations probe the inner regions of the protostar, where planets are expected to form and inherit the chemical composition of their natal environment, providing a direct link between ice sublimation and gas-phase chemistry. Our results represent the first high-spectral-resolution mid-infrared view of both water and methanol toward a low-mass protostar, offering a unique window into the chemical composition of the innermost envelope and planet-forming region, and highlighting the diagnostic power of high-resolution mid-infrared spectroscopy to uncover hidden chemical layers and the ice-to-gas transition in embedded protostars.

\end{abstract}

\keywords{Astrochemistry (75) --- Star formation (1569) --- Interstellar medium (847) --- Interstellar molecules (849) --- Chemical abundances (224) --- individual objects: SVS13-A}

\section{Introduction} \label{sec:intro}
Water and methanol are among the most abundant molecular components of interstellar ice mantles, playing key roles in the chemistry of star- and planet-forming regions. Water is expected to dominate the ice composition, forming efficiently on grain surfaces and, in warm dense regions, through neutral-neutral gas-phase reactions. Chemical models predict that in the innermost parts of protostellar envelopes, where temperatures exceed the sublimation threshold of icy mantles, water becomes highly abundant in the gas phase, reaching fractional abundances up to $\sim$10$^{-4}$ with respect to H$_2$ \citep[e.g.,][]{Ceccarelli1996, van_dishoeck_water_2014}. Similarly, methanol (CH$_3$OH) is formed almost exclusively on dust grains via successive hydrogenation of CO \citep{watanabe2002, rimola2014}, and is the most abundant interstellar complex organic molecule (iCOM\footnote{iCOMs are defined as molecules with $\ge$ 6 atoms which contain at least one C atom and another heavy element, such as O or N. \citep{herbst_complex_2009,ceccarelli_seeds_2017}}) in both the solid and gas phases.

Measuring the water content in astronomical sources is a very difficult task, because of the terrestrial atmosphere.
The saga only started in the late 90s with \textit{ISO}, and continued with \textit{SWAS}, \textit{ODIN}, \textit{Spitzer} and \textit{Herschel}.
All these instruments showed that water in star forming regions is very abundant ($10^{-6}-10^{-4}$ wrt H$_2$), containing almost all the elemental oxygen not locked into the refractory interstellar dust  grains or into CO and CO$_2$ \citep[e.g.,][]{ceccarelli_structure_2000, van_dishoeck_water_2014, boogert_observations_2015}.
However, it is largely frozen into the dust icy mantles \citep[e.g.,][]{boogert_observations_2015, mcclure2023}, except in the warm regions where the water ices sublimate \citep{ceccarelli_structure_2000,helmich_detection_1996,coutens_water_2014} or in shocked regions where they are sputtered \citep{liseau_thermal_1996,snell_water_2000, nisini_water_2010}.
Because of their limited spatial resolution, these observations mainly probed the large-scale environment of the observed sources.
In contrast, methanol transitions at millimeter wavelengths are more accessible and have been used extensively to trace both physical and chemical structure in protostellar systems \citep[e.g.,][]{cazaux2003, Walsh_2016, de_simone_hot_2020,Ilee_2026}.

In Solar-type protostars, water-ice-sublimated regions are called {\it hot corinos} \citep{ceccarelli_hot_2004}.
They have dust temperatures $\gtrsim$ 100 K, densities $\gtrsim$ 10$^7$ cm$^{-3}$ and size $\lesssim$ 100 au, and are found around Class 0/I \citep[10$^4$-10$^5$ yr;][]{andre_prestellar_2000} protostars \cite[e.g.,][and references therein]{bianchi_astrochemistry_2019, de_simone_hot_2020, belloche_questioning_2020}.
Recent studies suggest that signs of planet formation may already be present at these early stages of protostellar evolution, as indicated by the detection of gaps and rings \citep[e.g.,][]{fedele_alma_2018, sheehan_multiple_2018} and by constraints from the elemental mass budget \citep[e.g.,][]{tychoniec_dust_2020}. 
Therefore, characterizing chemically the young protostellar system is crucial to understand the chemical content a forming planet can inherit \citep{bianchi_census_2019,Drozdovskaya_2019}. 

We conducted high spectral resolution (R=70,500)
spectroscopy from 26 $\mu$m with EXES \citep[Echelon Cross Echelle Spectrograph][]{richter2018} on board of SOFIA toward the chemically rich Class I hot corino SVS13-A in the Perseus/NGC 1333 star forming region, at a distance of $\sim$300 pc \citep{zucker_mapping_2018}. 
As a reference, high spectral resolution MIR observations have already provided insight into the characteristics of massive protostars revealing rich absorption spectra \citep[e.g.,][]{knez2009,Indriolo2015, Indriolo2018, Indriolo2020, Dungee2018, Rangwala2018, Nickerson2021ApJ, Nickerson2023, Barr2018, Barr2020, Barr2022, Li_2023ApJ...953..103L}. 

SVS13-A is a well-studied low-mass Class I protobinary system \citep[L$_{submm}$/L$_{bol}$ $\sim$ 0.008;][]{chen_iram-pdbi_2009} and possesses a hot corino region, where several iCOMs are detected and imaged \citep{de_simone_glycolaldehyde_2017,bianchi_census_2019, bianchi_two_2022A, diaz-rodriguez_physical_2022}.  
Millimeter observations revealed the presence of both methanol and water: single-dish analysis of HDO lines provided evidence for warm ($>$150 K) and dense gas in a compact (below 50 au) region \citep{codella_hot_2016}, while interferometric imaging with ALMA suggests that HDO traces an accretion shock at the interface between a streamer and the disk \citep{bianchi_streamers_2023}. Similarly, CH$_3$OH lines observed with ALMA and IRAM-30m indicate column densities up to 10$^{19}$ cm$^{-2}$ and excitation temperatures of 100--200 K, consistent with emission from the hot corino \citep{bianchi_census_2019, bianchi_two_2022A}.

\section{Observations} \label{sec:observations}
SVS13-A was observed on Feb. 25, Mar. 08, and Mar. 11, 2022 (UT), with EXES on SOFIA in the cross-dispersed, high-low mode, centered at 26.26 $\micron$ under program ID 09-0162 (PI: De Simone). The data can be downloaded from the SOFIA archive hosted at IRSA \citep{https://doi.org/10.26131/irsa636}. The spatial resolution for the observation was $\sim$2.6$''$, approximately the diffraction limit  \citep{exeshandbook2022}. 

The slit length and width were 11$''$ and 3.2$''$, respectively. We derived a spectral resolution R=70500 $\pm$ 2000 ($\sim$ 4.25 km s$^{-1}$) for the data, which was calculated by fitting atmospheric transmission models to narrow telluric lines captured in this data set and archival data taken in the same mode at nearby wavelengths.  
SVS13-A was nodded along the slit by 5$''$ at 1 minute intervals to facilitate sky and thermal background subtraction. Spectral flat fields were taken before each observation to remove the order blaze efficiency shape and pixel response variations.
In total, the on-source integration time was 125 minutes (59.7 min, 27.2 min and 38.4 min on each date, respectively).

The data were reduced using the Python version of SOFIA Redux \citep{clarke2015}. The standard reduction sequence included noise spike removal, nod-subtraction, flat-field division, order rectification and coadding of the nod-subtracted pairs. The pipeline then optimally extracts the A and B traces in the 2-D echellogram, which are averaged together into a single 1-D spectrum. The wavelength solution was calculated using atmospheric lines measured in the radiance spectra, produced with the same data processed with the nod-subtraction step turned off. 
The wavelength calibration of the data was $\sim$ 0.3 km s$^{-1}$, based on our analysis of archival 20$-$28 $\mu$m EXES spectra (Program ID 75-0106, P.I. Montiel), observed at very high S/N in the same configuration as the SVS13-A data set. 

Two custom procedures were used in place of the standard pipeline procedures in order to improve the S/N of the spectra. First, in some A-B nod pairs, there was bright spurious signal near the order edges caused by mechanical shifts in the internal optics that occurred during the integration. We corrected these frames at the raw data step by shifting the A-side nod integrations with respect to the B-side until the residual bright edges were minimized. 
Secondly, we used a custom routine for the outlier-resistant weighted mean of the nod-pairs. The routine forces the total flux per wavelength channel in each A-B nod pair to be zero, assuming that SVS 13A is observed with the same flux on both sides of the nod. This step cancels residual telluric emission and improves the outlier removal when the nod-pair frames are averaged together. 
The spectra from each observation were corrected for telluric transmission and instrumental baseline slopes near the lines of interest. The details of this process are described in Appendix \ref{sec:telluriccorrection}. We combined the transmission corrected spectra by shifting the wavelength scale of each observation to the local standard of rest (LSR) frame and interpolating the flux and flux errors to a common wavelength array. The weighted mean of the fluxes in each wavelength channel was calculated using 1/$\sigma_{f}^2$ weights, where $\sigma_{f}$ is the uncertainty in the normalized flux.  The typical signal-to-noise ratio in the combined spectrum was $\sim$140 per resolution element. 

\section{Results and Analysis} \label{sec:analysis}
Seven methanol lines and two water lines were detected in absorption ($\ge 4\sigma$). The upper limit on a third water line provided useful leverage on the gas temperature. The spectroscopic parameters of the lines are listed in Table \ref{tab:moldetections}. 

We performed a preliminary analysis using Gaussian line fits and rotation diagrams, but we found that the uncertainties in the derived physical parameters could be improved by a factor of $\sim$2 by applying LTE slab model fitting. The following section describes the slab model fitting procedure and results. The rotational diagram analysis is presented in Appendix \ref{sec:rotationaldiagrams}. 

The spectroscopic line data for H$_2$O and $\nu_{12}$ CH$_3$OH was taken from the HITRAN 2020 database \citep{gordon2022} and from \citet{nickerson_discovery_2025}, respectively. The latter line list was introduced with the first astrophysical detection of $\nu_{12}$ CH$_3$OH toward the massive protostar NGC 7538 IRS 1.

The rotation diagram analysis and slab model fits both begin with the same set of equations and assumptions. We assume that the absorbing gas is in local thermodynamic equilibrium (LTE), and the level populations can be described by the Boltzmann distribution \citep{goldsmith_population_1999}:
\begin{equation}
    \frac{N_l}{g_l} = \frac{N_\textrm{tot}}{Q(T_\textrm{ex})} \textrm{exp}\left(-\frac{E_l}{T_\textrm{ex}}\right), \label{eq:boltz}
\end{equation}
where $E_l$ is the lower level energy of the transition expressed in Kelvin, $Q(T_\textrm{ex})$ is the partition function, and the $T_\textrm{ex}$ and $N_\textrm{tot}$ are the excitation temperature and the total column density of the molecule traced by the absorption lines.  

We computed level-specific column densities using Eq. \ref{eq:N_l}, which assumes that the absorption lines are optically thin and stimulated emission can be neglected \citep{mangum2015}. 

\begin{equation}
N_l = \frac{8\pi}{A_{ul} \lambda^3} \frac{g_l}{g_u} \int \tau(v)dv, \label{eq:N_l}
\end{equation}

where $\lambda$ is the line wavelength, $A_{ul}$ is the spontaneous emission coefficient, $g_l$ and $g_u$ are the statistical weights of the lower and upper levels, $v$ the velocity, and $\tau(v)$ is the optical depth. 

We constructed our slab models using Eq. \ref{eq:N_l} and \ref{eq:boltz}, assuming Gaussian line profiles, which allows the the substitution, $\int \tau(v)dv = \frac{b\tau_0}{\sqrt{\pi}}$ in Eq. \ref{eq:N_l}, where b is the Doppler parameter in km s$^{-1}$ ($FWHM = 2\sqrt{2}b$). 
In combination with \ref{eq:boltz}, the formula for peak line opacity becomes:

\begin{equation}
    \tau_0 = \frac{10^{-17}A_{ul}g_u\lambda^3}{8 \pi^{3/2} b}\frac{N_{tot}}{Q(T_{ex})} 
 \textrm{exp}(-E_l/T_{ex}), \label{eq:taumax}
\end{equation}
where $\lambda$ is the rest wavelength of the line transition in $\mu$m, and E$_l$ is its lower level energy in K.
The line opacity profile was calculated each line using its $\tau_0$ value, FWHM and $v_{LSR}$ and interpolated onto a uniform wavelength grid, to be summed with the opacities all the other lines included in the model. The simulated spectrum was smoothed to the instrumental resolution and the optical depth was converted to flux.

Using $\chi^2$ to represent the fit quality, we calculated the posterior distributions to determine physical parameters and uncertainties for T$_{rot}$, N$_{tot}$, FWHM and v$_{LSR}$ using the Markov Chain Monte Carlo (MCMC) sampler \textit{mcmc} \citep{foreman-mackey2013} with linear sampling over the parameter ranges: $50 < T_{ex} < 250 $ K, $14 <\log{N_{tot}} < 19$ cm$^{-2}$, $0.1 < $ FWHM$ < 10 $ km s$^{-1}$, $0 < $ v$_{LSR} < $ 20 km s$^{-1}$. We used the 16$^{th}$ and 84$^{th}$ percentiles of the likelihood distribution for our parameter uncertainties. 

\begin{figure*}[ht!]
    \includegraphics[width=\textwidth]{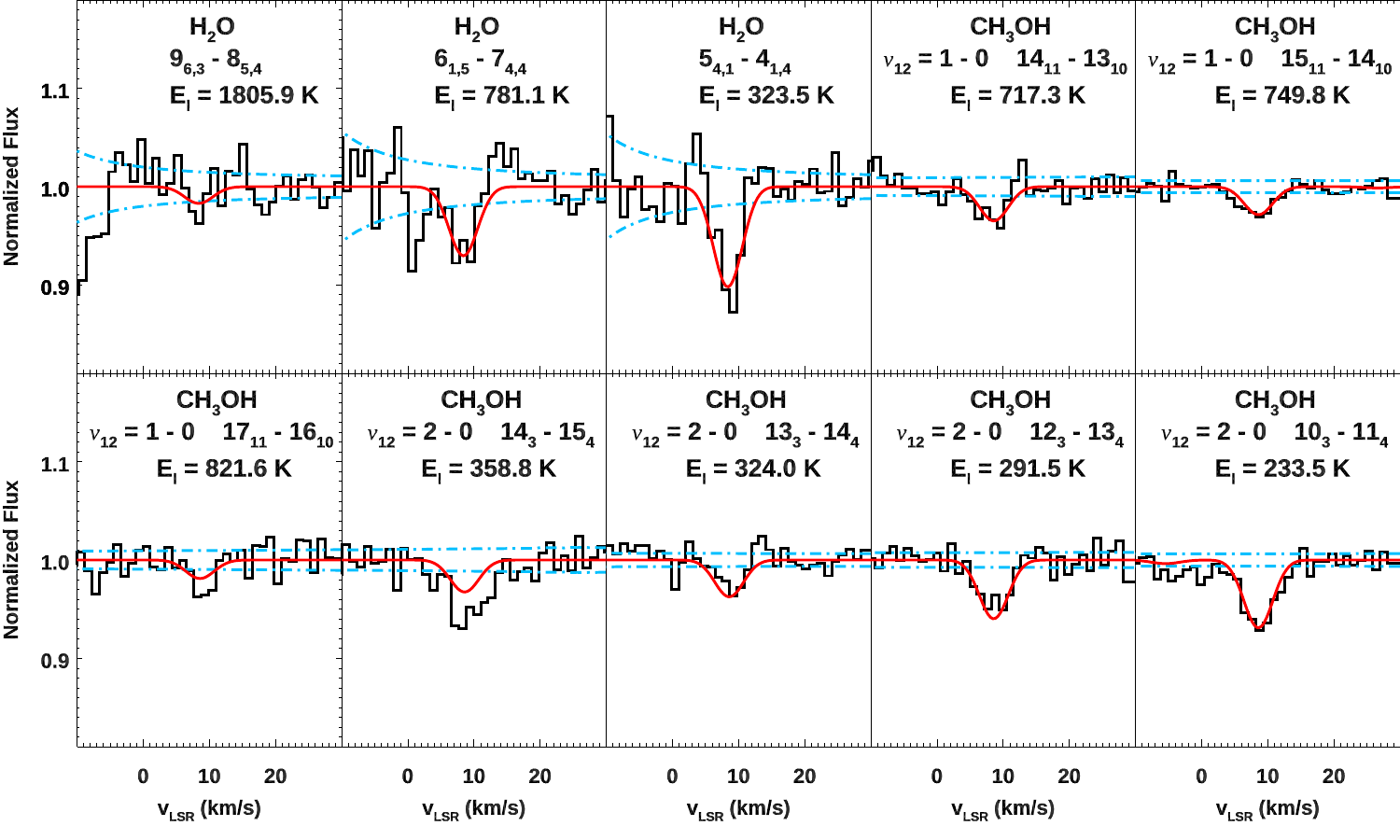}
    \caption{Observed H$_2$O and CH$_3$OH line profiles with the best fitting LTE slab models overplotted in red. The dashed blue line indicates the 1-$\sigma$ flux uncertainty.  In the H$_2$O profiles, the nearby atmospheric water absorption causes the flux uncertainty to rise quickly toward lower velocities. The spectra has been binned by 3 pixels.}
    \label{fig:waterandmethanol_profiles}
\end{figure*}

\begin{figure*}[ht!]
    \centering
    \includegraphics[width=\textwidth]{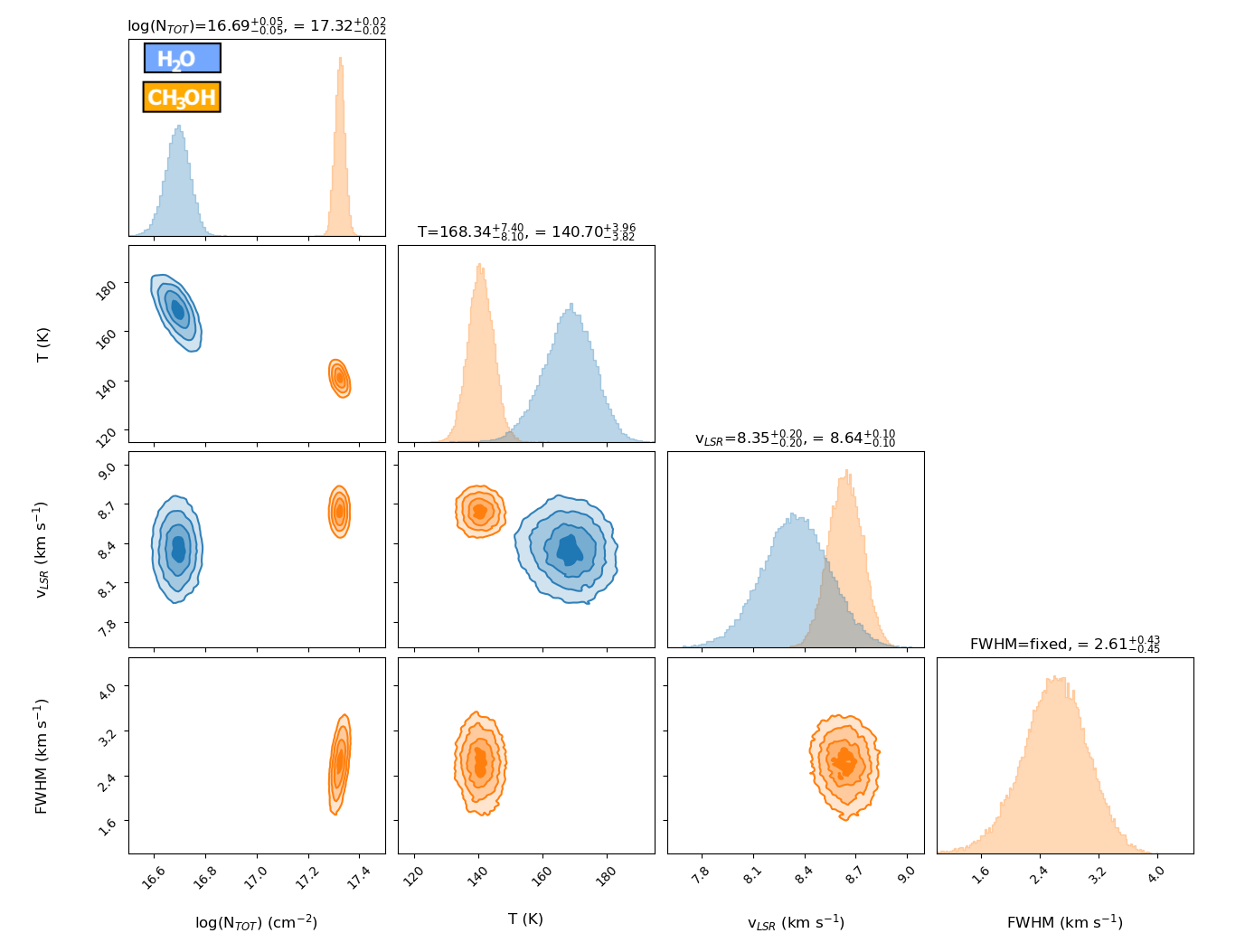}
    \caption{Posterior distributions for the slab model fits to water(blue) and methanol(orange). The best fitting parameter values are listed for water, followed by methanol, above each column. The line width of water was fixed to the value determined for methanol to avoid physically unrealistic values. Hence, the posterior likelihood distribution for the FWHM of water is not shown.} 
    \label{fig:water_and_methanol_corner}
\end{figure*}

Corner plots from the slab model fits to CH$_3$OH and H$_2$O are shown in Figure \ref{fig:water_and_methanol_corner}. Table \ref{tab:slabmodelparameters} lists the values for the physical parameters derived with the model fits.  In Figure \ref{fig:waterandmethanol_profiles}, we display the line profiles of both molecules with the best fitting slab models overplotted in red.

The velocity of the absorption was found to be $+8.4^{+0.2}_{-0.2}$ km s$^{-1}$ and $+8.6^{+0.1}_{-0.1}$ km s$^{-1}$ for water and methanol respectively, consistent with the systemic velocity of the protostellar system \citep[+8.5 km s$^{-1}$;][]{codella_hot_2016,bianchi_census_2019,bianchi_two_2022A}.

For CH$_3$OH we derived T$_{ex}$ = 141$^{+4}_{-4}$ K and N$_{tot}= (2.09^{+0.09}_{-0.09}) \times 10^{17}$ cm$^{-2}$. Note that the CH$_3$OH detections consist of both the $\nu_{12}$ fundamental and its overtone, but there are not enough lines to perform separate analysis so we have assumed the vibrational levels are thermalized. CH$_3$OH has A-type and E-type symmetry states that do not interact radiatively.
All our line detections had A-type symmetry and following \citet{bianchi_deuterated_2017} we applied the equilibrium ratio E/A $=$ 1 appropriate for warm gas. 

For H$_2$O, we derive T$_{ex}$ = 168$^{+7}_{-8}$ K and N$_{tot}= (4.9^{+0.6}_{-0.6}) \times 10^{16}$ cm$^{-2}$. We have assumed an ortho-to-para ratio of o/p = 3 for water, which is supported by observational studies of star forming regions in the far-infrared with Herschel Observatory \citep{vandishoeck2021_A&A...648A..24V}. 
The undetected H$_2$O line was found to have negligible impact on the best fit N$_{tot}$ and T$_{ex}$ values but its inclusion reduced the uncertainty of T$_{ex}$ by 20 $\%$. 

The corner plots for slab model fits to H$_2$O are shown in Figure \ref{fig:water_and_methanol_corner}. 
Like in the individual Gaussian line fits (see \ref{sec:rotationaldiagrams}), the slab models find widths that are narrower than the resolution limit, so we fixed the intrinsic line width to be 2.6 km s$^{-1}$ (5.0 km s$^{-1}$ at the instrumental resolution), adopted from the methanol slab analysis. We explored the impact of fixing the intrinsic line width of H$_2$O by repeating the MCMC fits using values between 0.1 and 3 km s$^{-1}$ and the results for the three free parameters were found to be insensitive to which linewidth value was adopted.
In Figure \ref{fig:waterandmethanol_profiles} we show the line profiles for the 3 targeted H$_2$O lines with the best fitting slab model, in red.  

Note that deriving absolute abundances is not straightforward. No measurement of warm H$_2$ is available in the infrared for the same gas traced by our absorption features, and the H$_2$ emission detected by \citet{hodapp2014} has been associated with the jet. One could adopt the H$_2$ column density derived from mm emission by \citet{bianchi_two_2022A}, but the IR absorption and mm emission likely trace different regions (see Section \ref{sec:discussion}). As a rough estimate, assuming an H$_2$ density of $\sim$10$^8$ cm$^{-3}$ and that the absorbing gas extends over $\sim$200 au, we would obtain $N$(H$_2$) $\sim$ 3$\times$10$^{23}$ cm$^{-2}$, implying a CH$_3$OH abundance of order 10$^{-6}$–10$^{-7}$. However, these values should be taken as indicative only.

The similarity in velocity and temperature indicates that the methanol and water lines probe the same location within the hot corino.  

\begin{table}[ht!]
\caption{LTE slab Model Fit Physical parameters. Column density (N$_{\rm tot}$), rotational temperature T$_{\rm rot}$, rest velocity ($v_{LSR}$) and width of the line (FWHM). }
\renewcommand{\arraystretch}{1.2}
\label{tab:slabmodelparameters}
\vskip0.03in
\begin{center}
\begin{tabular}{c|c|c}
\hline
\hline
 & H$_2$O & CH$_3$OH \\
\hline
T$_{\rm rot}$ & $168^{+7}_{-8}$ K & $141^{+4}_{-4}$ K \\
N$_{\rm tot}$ & $4.9^{+0.6}_{-0.6}$ $\times$ 10$^{16}$ cm$^{-2}$ & $2.09^{+0.09}_{-0.09}$ $\times$ 10$^{17}$ cm$^{-2}$ \\
$v_{LSR}$ & $+8.4^{+0.2}_{-0.2}$ km s$^{-1}$ & $+8.6^{+0.1}_{-0.1}$ km s$^{-1}$ \\
FWHM & 2.6\footnote{The H$_2$O line widths were fixed to the value from the CH$_3$OH LTE model fits.} km s$^{-1}$ & $2.6^{+0.4}_{-0.4}$ km s$^{-1}$ \\
\hline
\hline
\end{tabular}
\end{center}
\end{table}

\section{Discussion}\label{sec:discussion}
We observed SVS13-A in the mid-infrared with SOFIA at 26 \micron, where we detect several water and methanol lines in absorption. Due to the $\sim2.6''$ angular resolution, these observations cannot spatially resolve the two components of the close binary (VLA4A and VLA4B), separated by $\sim0\farcs3$ ($\sim$90 au at 300 pc). Historically, the identification of the MIR source has been debated. Based on earlier radio and optical data, \citet{anglada_discovery_2000} proposed that only VLA4B is surrounded by a dusty envelope or disk and dominates the millimeter emission, while VLA4A corresponds to the visible component. Later, \citet{hodapp2014, fujiyoshi2015}, identified VLA4B as the origin of the MIR emission using accurate 2MASS astrometry. 
Both components are associated with compact hot corinos, as confirmed by ALMA observations of iCOMs and deuterated water \citep{codella_hot_2016, de_simone_glycolaldehyde_2017, diaz-rodriguez_physical_2022, bianchi_two_2022A}. 

The methanol and water absorption lines peak at the protostellar velocity, suggesting an origin in warm gas near the central system.
Since our observations do not spatially resolve the binary, it is not straightforward to determine which component the absorption arises from. The line velocity is closer to that of VLA4B ($\sim$+8.5 km s$^{-1}$) than VLA4A ($\sim$+7.7 km s$^{-1}$) \citep[][]{bianchi_two_2022A, bianchi_streamers_2023}, suggesting that the gas is more likely associated with VLA4B. Note that, the spectral resolution of our data is larger than the velocity difference between the two sources, so this conclusion should be taken with caution. However, when coupled with the fact that the mid-infrared continuum is dominated by VLA4B, it becomes plausible that the absorption traces warm gas along the line of sight toward this source.

The best fit LTE analysis yields rotational temperatures of $\sim170$ K for water and $\sim$140 K for methanol. 

These temperatures are consistent with a gas layer within the ice thermal sublimation zone. 
However, it is important to stress that, in an absorption regime, the derived rotational temperatures do not directly correspond to the kinetic temperature of the gas, because it reflects the level populations under the influence of the background continuum.
Indeed, since absorption occurs only when the excitation temperature is below the continuum brightness temperature, the actual gas temperature could be lower than the derived rotational values. 

Interestingly, we find that the methanol column density is $\sim$4 times higher than that of water, contrary to expectations since water is generally more abundant in interstellar ices and warm gas (see \ref{sec:intro}). Indeed, if the gas-phase abundances were dominated by the sublimation of the entire bulk of the ice mantles, the observed ratio should reflect the ice composition. However, ice surveys consistently find CH$_3$OH/H$_2$O ratios of less than 10\% \citep{Parise2003, boogert_observations_2015, mcclure2023}, and even in the case of complete mantle sublimation, the gas-phase ratio would remain far from the value that we measure.

ALMA observations of SVS13-A provide complementary information of its hot corino content: CH$_3$OH shows column densities $\sim10^{19}$ cm$^{-2}$ \citep{bianchi_two_2022A} and HDO $\sim10^{17}$ cm$^{-2}$ \citep{codella_hot_2016}, which, with their derived [D]/[H] ratio of $\sim1.5\times10^{-2}$, implies H$_2$O $\sim10^{19}$ cm$^{-2}$. These millimeter observations suggest that similar CH$_3$OH and H$_2$O column densities are possible, although optical depth effects and the different regions probed may bias these measurements. Our SOFIA results are likely probing cooler layers along the line of sight compared to the warm, dense gas dominating the millimeter emission. The mid-IR absorption may sample the outer layers of the same hot corino structure, above the $\tau= 1$ dust surface (or photosphere) or it could be located at larger distances in the circumbinary material (e.g. \citet{diaz-rodriguez_physical_2022}) while millimeter emission integrates over deeper and warmer regions. 
In this sense, absorption and emission probe different regions within the protostellar environment, challenging a direct comparison of the inferred abundance ratios.\\

It is possible that the water column density derived from our data is underestimated: in the infrared regime, both absorption and emission can occur. When the gas density is far below the critical density to populate the transition upper level, collisions are inefficient and the excitation temperature is mainly set by the radiation field. Against the bright MIR continuum (T$_{bg}>$100 K), this typically results in absorption. However, if the density exceeds the critical density, the level population becomes thermalized and emission is present if the gas is hotter than the continuum. For the strongest water transition detected here ($5_{4,1}-4_{1,4}$), the critical density estimated from Einstein coefficients and collisional rates \citep[C$_{ul}$;][]{Dubernet_2009}, computed as $\sim {A_{ul}}/{C_{ul}}$, is of order 10$^{10}$ cm$^{-3}$. Such densities may be reached in the innermost regions of SVS13-A \citep{bianchi_two_2022A}, meaning that some water emission from compact hot gas could partially fill in the absorption, leading to an underestimate of the derived column density. \\ 
For the methanol transitions detected here, collisional coefficients are not available, but the A$_{ul}$ values compared to water suggest higher critical densities, making emission less likely for the detected CH$_3$OH lines. A more robust assessment would require detailed radiative transfer modeling that couples both absorption and emission processes.
It is also worth noting that photodissociation effects are probably minor, since the MIR-absorbing layer is embedded in the SVS13-A envelope and heavily dust-shielded. While selective UV destruction of water could in principle affect the CH$_3$OH/H$_2$O ratio \citep[e.g.][]{Heays2017}, this scenario seems unlikely for SVS13-A.

Even accounting for this potential bias, the measured CH$_3$OH/H$_2$O ratio remains unusually high, suggesting that additional chemical or physical effects may enhance methanol relative to water. In Figure \ref{fig:schematic} we show a schematic picture of SVS 13-A as observed in the mid-infrared, where the absorbing molecular gas is seen primarily along the line of sight to VLA 4B. Due to the higher dust opacity in the MIR, the innermost layers of the corino that dominate the millimeter emission are not detected, instead probing the gas in front of the continuum in a volume that is apparently sublimating methanol more efficiently than water.
Two possible contributing factors for the selective desorption are discussed below:

\begin{figure*}[ht!]
    \centering
    \includegraphics[width=\textwidth]{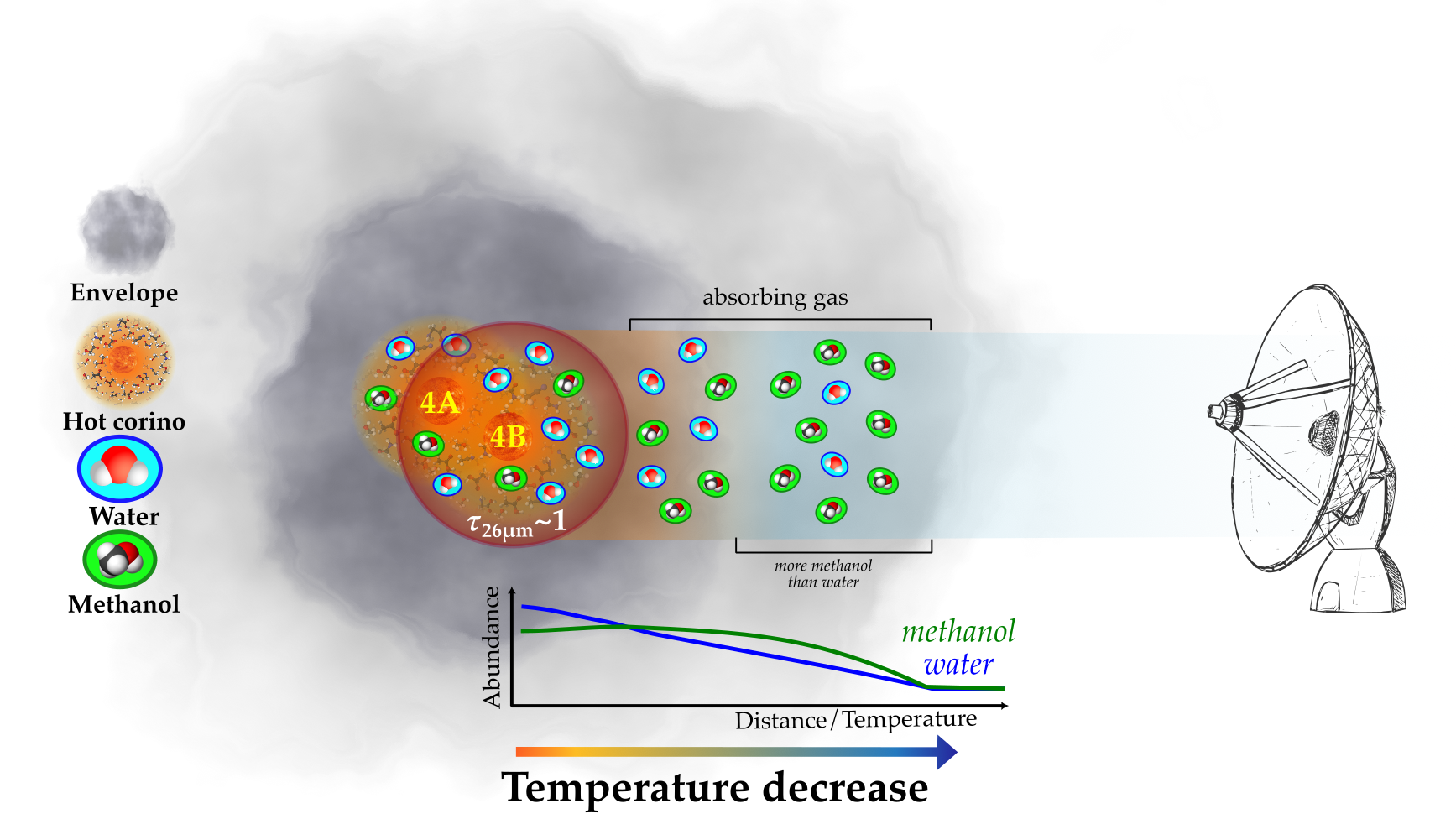}
    \caption{Artistic scheme of the VLA 4A/4B binary system, not in scale. Methanol and water absorption features originate from gas located in front of the dust photosphere at 26 $\mu$m, which is likely driven by VLA 4B (see text). The photosphere size and location is not certain, but the absorption selectively probes only a layer of sublimated ice, rather than the full bulk envelope, in a location where methanol may desorb more efficiently than water.}
    \label{fig:schematic}
\end{figure*}

- \textit{Binding energy influence:}
Recent quantum chemical calculations have demonstrated that molecules on dust grain surfaces do not desorb at a single, well-defined binding energy, but instead exhibit a distribution of binding energies that affects both their sublimation and freeze-out behavior. In particular, \cite{Tinacci_2023} and \cite{bariosco2025} have computed the binding energy distributions for H$_2$O and CH$_3$OH on water ices, respectively. While the peak of the distributions for the two species occurs at similar energies, the width and shape of the distributions differ. As a result, at temperatures below about 60 K, methanol can start to thermally desorb more efficiently than water. Assuming similar initial abundances in the ice mantles, this effect could lead to a differential desorption, with methanol being present in the gas phase in greater quantities than water by up to a factor of $\sim$2, when considering only the binding energies and not how the desorption exactly occurs. Additionally, once in the gas phase, water would deplete faster than methanol, accounting for an additional factor of 1.5. 

Recent laboratory and observational studies also indicate that the ice composition can influence the effective binding energies of frozen molecules. Observations of CO$_2$ ice bands toward massive protostars show that thermal processing can lead to ice segregation and to the formation of CO$_2$ and CH$_3$OH rich phases \citep[e.g.,][]{dartois1999}. In such cases, the ice may no longer be water-dominated, affecting molecular binding energies. Indeed, methanol in CO$_2$ ice shows a lower effective binding energy than in water ices \citep{dartois2020}, and CO-rich ices show similar effects \citep{molpeceres2024},  modifying their desorption behavior. However, the most recent theoretical calculations that explicitly include distributions of binding energies for methanol and water have been performed for H$_2$O dominated ice surfaces. These calculations therefore provide the most consistent framework currently available to compare the desorption behavior of H$_2$O and CH$_3$OH.

- \textit{Ice stratification:} Interstellar ice mantles may exhibit layered or mixed structures, which can significantly influence the thermal desorption behavior of embedded molecules. For example, \citet{Kruczkiewicz2024} showed that CH$_3$OH is only modestly trapped in compact amorphous water ice, and its desorption profile differs from that of other volatiles such as CO or NH$_3$. This can suggest that in a stratified ice mantle architecture, CH$_3$OH-rich regions may begin to desorb at lower temperatures than the underlying or co-existing H$_2$O-rich layers. 
Together with the effects of ice composition described above, this suggests that both segregation and stratification could enhance the gas-phase abundance of methanol relative to water.
However, with the information currently available, it is difficult to quantify how much of the observed abundance difference could be attributed to differences in the sublimated layers. In any case, such an interpretation would rely on the assumption that we are probing only a limited fraction of the sublimated ice mantle and not the whole bulk.

These high-resolution MIR absorption observations offer a unique window into the warm gas in the immediate vicinity of a low-mass protostar. Our SOFIA data provide the first high-resolution view of both water and methanol in the MIR, allowing us to probe the physical and chemical conditions of the circumstellar gas on scales comparable to the inner envelope or disk. This offers a novel perspective that was previously inaccessible.

\section{Conclusions}
We conducted high spectral resolution (R= 70,500; 4.25 km s$^{-1}$) spectroscopy at about 26 \micron \ with EXES on board SOFIA toward the hot corino region associated with the SVS13-A protostellar system. 
Several absorption lines of H$_2$O and CH$_3$OH are detected near the systemic velocity, tracing warm gas (T$\mathrm{rot}\sim150-170$ K) likely located in front of the mid-infrared continuum source. Remarkably, the CH$_3$OH column density is larger than that of H$_2$O by a factor of $\sim$4, challenging current expectations from interstellar ice compositions. This difference may arise because absorption selectively probes only a layer of sublimated ice, rather than the full bulk mantle. In this perspective, the observed column density ratio could reflect a combination of differential desorption efficiencies, layered or mixed ice structures, and physical stratification within the inner envelope. These findings highlight the unique diagnostic potential of high resolution MIR absorption spectroscopy to access otherwise hidden chemical layers in embedded protostars, offering new insights into the ice-to-gas transition in planet-forming environments. \\

\textit{Acknowledgments:} We are very grateful to the SOFIA team who supported  observations.
This project has received funding from: 
1) Italian Ministry for Universities and Research under the Italian Science Fund (FIS 2 Call - Ministerial Decree No. 1236 of 1 August 2023).
2) the European Research Council (ERC) under the European Union's Horizon 2020 research and innovation program, for the Project “The Dawn of Organic Chemistry” (DOC), grant agreement No 741002; 
3) the PRIN-MUR 2020  BEYOND-2p (Astrochemistry beyond the second period elements, Prot. 2020AFB3FX)
4) the project ASI-Astrobiologia 2023 MIGLIORA (Modeling Chemical Complexity, F83C23000800005)
5) the INAF-GO 2024 fundings ICES, and the INAF-GO 2023 fundings PROTO-SKA (Exploiting ALMA data to study planet forming disks: preparing the advent of SKA, C13C23000770005)
6) the European Union’s Horizon 2020 research and innovation programs under projects “Astro-Chemistry Origins” (ACO), Grant No 811312; 
7) the German Research Foundation (DFG) as part of the Excellence Strategy of the federal and state governments - EXC 2094 - 390783311. \\
Portions of this research were performed at the Jet Propulsion Laboratory, California Institute of Technology, under contract with the National Aeronautics and Space Administration and California Institute of Technology. Based on observations made with the NASA/DLR Stratospheric Observatory for Infrared Astronomy (SOFIA). SOFIA was jointly operated by the Universities Space Research Association, Inc. (USRA), under NASA contract NNA17BF53C, and the Deutsches SOFIA Institut (DSI) under DLR contract 50 OK 2002 to the University of Stuttgart. 

\vspace{5mm}
\facilities{SOFIA(EXES)}

\appendix
\section{Telluric correction and baseline correction}\label{sec:telluriccorrection}
Despite observing from the high altitudes reached by SOFIA, each source water line fell within the broad wings of its terrestrial counterpart, reducing the transmission to 50--70\%. The precise transmission value depends sensitively on the water vapor overburden along the flight path and must be determined individually for each observation. In contrast, the source methanol lines are mostly free from telluric interference, with atmospheric transmission exceeding 99\%. 

\begin{figure*}[ht!]
    \centering
    \includegraphics[width=0.95\textwidth]{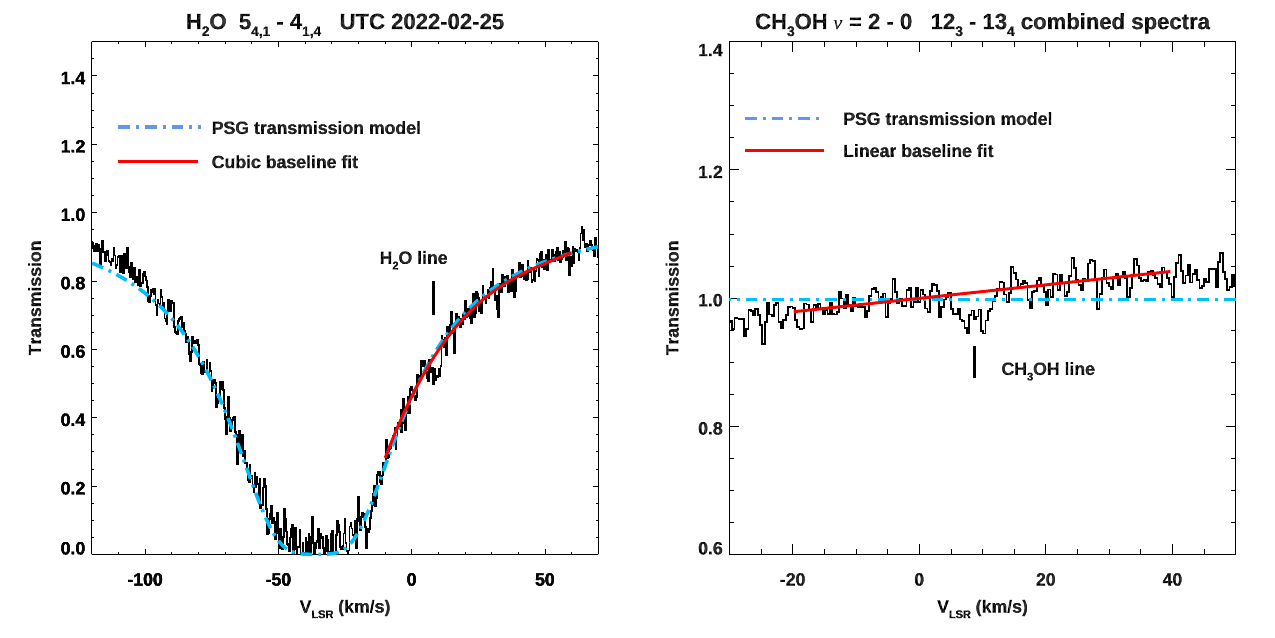}
    \caption{(Left) Example of the best fit PSG transmission model and the cubic polynomial used to refine the baseline over the $5_{4,1}-4_{1,4}$ source water line. The data is from the first night's observation on UT 2022-02-24. The source absorption line is visible at v$_{LSR}$= +8.5 km s$^{-1}$. (Right) Example of the linear baseline fit (red) for the $\nu_{12}=2-0$ $12_3-13_4$ methanol line. There is no significant atmospheric interference at this wavelength, as shown by the transmission model denoted by the dashed blue line.}
    \label{fig:baseline}
\end{figure*}

We generated initial telluric transmission models for each observation with Planetary Spectrum Generator (PSG) \citep{villanueva2018} using the average values of key observing parameters: date, time, latitude, longitude, altitude and zenith angle. The models were smoothed to the R = 70500 resolution of the data. 
The default models overestimated the atmospheric water vapor column near the three targeted water lines (listed in Table \ref{tab:moldetections}). To adjust the column, we applied an exponent between 0.5 - 1.0 to the transmission model until the best visual match was made to the atmospheric line shape. 
After division of the telluric transmission models, the regions around the water lines displayed residual slopes in the continuum due to instrumental fringing or incomplete telluric correction.  We fixed the value of the transmission near the source water lines the derived PSG model and refit the region with a cubic polynomial using continuum points at -20 $<$ v$_{LSR}$ $<$ 4 and 14 $<$ v$_{LSR}$ $<$ 60 km s$^{-1}$. The velocity limits were chosen to avoid poorly corrected regions at  v$_{LSR} <$ -20 km s$^{-1}$ and contain enough well-behaved parts of the continuum to give confidence in the baseline shape. 
The cubic polynomial was divided from the uncorrected data, simultaneously correcting for the transmission and the instrumental baseline near each H$_2$O line. The left panel of Figure \ref{fig:baseline} shows an example of the PSG atmospheric model fit to the $5_{4,1}-4_{1,4}$ line on the first night's observation and the cubic polynomial used to refine the baseline near over the line location. 

The methanol lines had negligible interference from the atmosphere. For these lines, we made our baseline corrections after the spectra was combined by fitting first order polynomials to the continuum at -20 $<v_{LSR}<$ -4 km s$^{-1}$ and 14 $<v_{LSR} <$ 40 km s$^{-1}$. In the right panel of Figure \ref{fig:baseline}, we show an example of the the baseline fit near a methanol line in the combined spectrum. 

\section{Individual Line Fits and Rotation Diagram Analysis}\label{sec:rotationaldiagrams}

The tellurically corrected and combined spectra were converted to optical depth, via $\tau(\lambda) = -\textrm{ln}(F_\lambda/F_c)$, where $F_\lambda$ is the observed flux within a given wavelength bin and $F_c$ is the local continuum flux.
We fit a single component Gaussian function of the form $\tau(v)=\tau_0\exp(\frac{(v-v_{LSR})^2}{b^2})$ to line locations, where $\tau_0$ is the peak optical depth, $v_{LSR}$ is the velocity centroid in the LSR frame, and $b$ is the Doppler parameter and is related to full width at half maximum via FWHM$= 2\sqrt{\ln{2}}$b. The line transition data and Gaussian fitting results are listed in Table \ref{tab:moldetections}. Detections of the methanol lines ranged in significance between $\sim5-17\sigma$, where we define significance as the equivalent width (E.W.) divided by the E.W. uncertainty, measured over the interval $v_{LSR}$ $\pm$ $\frac{FWHM}{2}$ km s$^{-1}$. The H$_2$O $5_{4,1}-4_{1,4}$ and $7_{4,4}-6_{1,5}$ lines were detected with $\sim8\sigma$ and $\sim5\sigma$ significance. The H$_2$O $9_{6,3}-8_{5,4}$ line was not detected (1.7$\sigma$) but its upper limit was a useful constraint on the population in the slab modeling analysis. 

The weighted mean velocity and width of the methanol lines was +8.6 $\pm$ 0.2 km s$^{-1}$ and FWHM = 5.0 $\pm$ 0.2 km s$^{-1}$. The intrinsic line width is FWHM $\approx$ 2.7 km s$^{-1}$ after subtracting the instrumental resolution in quadrature. 

To derive the column density and temperature we use rotational diagram analysis following equations from \cite{mangum2015}. Assuming that the absorption lines are optically thin and that stimulated emission can be neglected, we compute the the level-specific column densities using equation \ref{eq:N_l}.
All line detections are measured with $\tau_0 < $0.15, which supports the optically thin assumption. 

By plotting $\ln(N_l/g_l)$ versus $E_l$ for the absorption lines and performing a linear fit, we constructed a rotational diagram and derive $T_\textrm{ex}$ and $N_\textrm{tot}$ from the slope and intercept of the fitted line, respectively. For methanol we derive $T_\textrm{ex}$ = 144$\pm$ 6 K and $N_\textrm{tot}$ = (2.5 $\pm$ 0.3) $\times$ 10$^{17}$ cm$^{-2}$.
 
\begin{figure*}[ht!]
    \centering
    \includegraphics[width=0.95\textwidth]{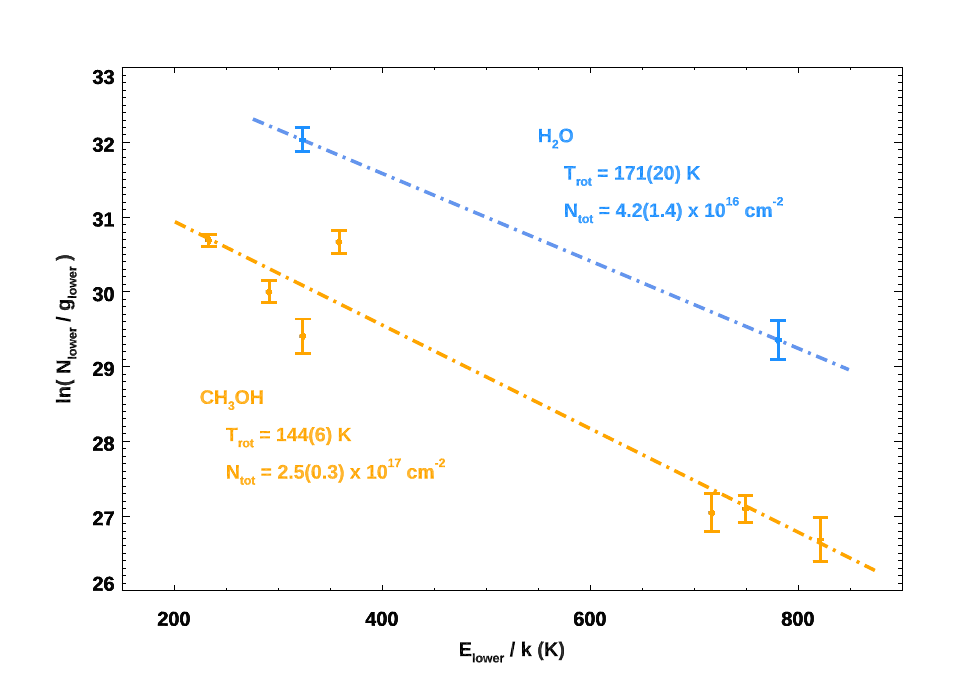}
    \caption{Rotation Diagram of the individually detected CH$_3$OH and H$_2$O lines toward SVS13-A. } 
    \label{fig:rotdiag}
\end{figure*}

Gaussian fits to H$_2$O $5_{4,1}-4_{1,4}$ and $7_{4,4}-6_{1,5}$ lines yielded widths that were nominally below the resolution limit: FWHM = 3.2 $\pm$ 0.4 km s$^{-1}$ and FWHM = 3.7 $\pm$ 0.8 km s$^{-1}$, respectively. We attribute this to noise fluctuations and note that the unconstrained Gaussian fits of two CH$_3$OH were also below the resolution limit, but the weighted mean FWHM was physically permissible (5.0 km s$^{-1}$).  
Figure \ref{fig:rotdiag} shows the rotation diagram for both molecules, with the derived excitation temperatures and column densities indicated. The results for all physical parameters are fully consistent with the slab model analysis. 

\begin{table*}[ht!]
\begin{center}
\caption{Spectroscopic line parameters}
\renewcommand{\arraystretch}{1.1}
\label{tab:moldetections}
\begin{tabular}{l|c|c|c|c|c|c|c|c|c|c}
\hline
\ \ \ \ H$_2$O Transition  & $\lambda$ & A$_{ul}$ & E${_{l}}$ & g$_{u}$ & g$_{l}$ & $\tau_0$ & v$_{LSR}$ & FWHM & N$_l$ & Signif.\protect\footnote{Significance refers to the ratio of the Equivalent width (E.W.) to the E.W. uncertainty between the range $v_{LSR}$ $\pm$ FWHM/2, where FWHM is the observed linewidth of CH$_3$OH, 5.1 km s$^{-1}$.}   \\
$J'_{Ka',Kc'}$ -- $J''_{Ka'',Kc''}$ isomer & ($\mu$m) & (s$^{-1}$) & (K) &  &  &  & (km s$^{-1}$) & (km s$^{-1}$) & (cm$^{-2}$) & \\
\hline
$9_{6,3}$ -- $8_{5,4}$ ortho & 26.580456 & 15.42 & 1805.9 & 57 & 51 & 0.048(0.015)& +7.9(0.4)& 1.9(0.8) & 7.4(3.8) x 10$^{11}$  & 1.7$\sigma$ \\
$7_{4,4}$ -- $6_{1,5}$ para  & 25.984951 & 0.597 & 781.1 & 15 & 13 & 0.090(0.014) & +8.4(0.3) & 3.7(0.8)  & 7.3(1.9) x 10$^{13}$  & 4.6$\sigma$ \\
$5_{4,1}$ -- $4_{1,4}$ ortho & 25.940150 & 0.026 & 323.5 & 33 & 27 & 0.147(0.015) & +8.7(0.2) & 3.2(0.4)  & 2.2(0.4) x 10$^{15}$  & 7.9$\sigma$ \\
\hline 
\ \ \ \ CH$_3$OH Transition & $\lambda$ & A$_{ul}$ & E${_{l}}$ & g$_{u}$ & g$_{l}$ & $\tau_0$ & v$_{LSR}$ & FWHM & N$_l$ & Signif.  \\
$v'$ -- $v''$\hspace{0.2cm} $J'_{K'}$ -- $J''_{K''}$\hspace{0.2cm}Sym. & ($\mu$m) & (s$^{-1}$) & (K) &  &  &  & (km s$^{-1}$) & (km s$^{-1}$) & (cm$^{-2}$) & \\
\hline 
$\nu_{12}=$ 1 -- 0 \hspace{0.2cm}$14_{11}$ -- $13_{10}$ A & 26.540842& 1.484 & 717.3 & 116 & 108 &  0.038(0.006)& +8.0(0.4) & 4.4(0.9) & 1.5(0.4) x 10$^{13}$ & 5.1$\sigma$ \\
$\nu_{12}=$ 1 -- 0 \hspace{0.2cm}$15_{11}$ -- $14_{10}$ A & 26.432589& 1.443 & 749.8 & 124 & 116 &  0.031(0.004)& +8.2(0.3) & 5.8(0.8) & 1.7(0.3) x 10$^{13}$ & 7.2$\sigma$ \\
$\nu_{12}=$ 1 -- 0 \hspace{0.2cm}$17_{11}$ -- $16_{10}$ A & 26.219625& 1.433 & 821.6 & 140 & 132 &  0.041(0.008)& +9.0(0.3) & 3.2(0.7) & 1.3(0.4) x 10$^{13}$ & 4.7$\sigma$ \\
$\nu_{12}=$ 2 -- 0 \hspace{0.2cm}$14_{3}$ -- $15_{4}$ A & 26.544862& 0.109  & 358.8  & 116 & 124 &  0.062(0.006)& +9.1(0.4) & 7.4(0.8) & 6.5(1.0) x 10$^{14}$ & 8.8$\sigma$ \\
$\nu_{12}=$ 2 -- 0 \hspace{0.2cm}$13_{3}$ -- $14_{4}$ A & 26.425527& 0.107  & 324.0  & 108 & 116 &  0.039(0.005)& +9.1(0.2) & 3.0(0.6) & 1.7(0.4) x 10$^{14}$ & 6.7$\sigma$ \\
$\nu_{12}=$ 2 -- 0 \hspace{0.2cm}$12_{3}$ -- $13_{4}$ A & 26.307780& 0.150  & 291.5  & 100 & 108 &  0.051(0.005)& +8.5(0.3) & 5.3(0.6) & 2.9(0.4) x 10$^{14}$ & 10.1$\sigma$ \\
$\nu_{12}=$ 2 -- 0 \hspace{0.2cm}$10_{3}$ -- $11_{4}$ A & 26.076765& 0.141  & 233.5  & 84  & 92  &  0.073(0.004)& +8.7(0.2) & 5.7(0.4) & 4.9(0.4) x 10$^{14}$ & 17.2$\sigma$ \\
\hline
\hline
\end{tabular}
\end{center}
\end{table*}


\bibliography{svs13_ALL}{}
\bibliographystyle{aasjournal}



\end{document}